\documentclass{article}

\usepackage{amsmath,amsfonts,amssymb,natbib,xcolor,graphicx,hyperref,cleveref,amsthm}
\usepackage{textcomp}

\usepackage[margin=1in]{geometry}

\newcommand{\R}{\mathbb{R}}
\newcommand{\E}{\mathbb{E}}

\renewcommand{\P}{\mathbb{P}}
\newcommand{\PP}{\mathbb{P}}
\renewcommand{\d}{\mathrm{d}}

\newcommand{\mY}{\mathcal{Y}} 
\newcommand{\mS}{\mathcal{S}}

\newcommand{\x}{x}
\newcommand{\no}{}
\newcommand{\N}{n}
\newcommand{\f}{f}
\newcommand{\lo}{\text{lo}}
\newcommand{\hi}{\text{hi}}

\crefname{prop}{proposition}{propositions}
\Crefname{prop}{Proposition}{Propositions}
\crefname{asp}{assumption}{assumptions}
\Crefname{asp}{Assumption}{Assumptions}
\crefname{thm}{theorem}{theorems}
\Crefname{thm}{Theorem}{Theorems}
\crefname{lem}{lemma}{lemmas}
\Crefname{lem}{Lemma}{Lemmas}
\crefname{cor}{corollary}{corollaries}
\Crefname{cor}{Corollary}{Corollaries}
\crefname{rem}{remark}{remarks}
\Crefname{rem}{Remark}{Remarks}

\newtheorem{prop}{Proposition}
\newtheorem{asp}{Assumption}
\newtheorem{thm}{Theorem}
\newtheorem{lem}{Lemma}

\newtheorem{rem}{Remark}

\begin{document}

\title{A foundation for the distance sampling methodology}

\author{B.R. Baer, L. Thomas, and S.T. Buckland \\ Centre for Research into Environmental and Ecological Modelling, \\ School of Mathematics and Statistics, University of St Andrews, \\ The Observatory, Buchanan Gardens, St Andrews, KY16 9LZ}

\maketitle

\begin{abstract}
    The population size (``abundance'') of wildlife species has central interest in ecological research and management. 
    Distance sampling is a dominant approach to the estimation of wildlife abundance for many vertebrate species. 
    Although distance sampling has been successfully applied and developed for decades, its statistical foundation is not complete: there are published proofs and arguments highlighting deficiency of the methodology. 
    This work provides a statistical foundation for distance sampling that has attainable assumptions. 
    In addition, we show that the identification and consistency of the developed distance sampling abundance estimator is unaffected by detection heterogeneity, 
    resolving a conjecture made over four decades ago. 
\end{abstract}

\section{Introduction}

One of the most basic questions concerning a wildlife population is: how many are there? 
Researchers in the field of statistical ecology have developed or adapted a suite of survey and inference methods to answer this question \citep{king2014statistical}. 
Distance sampling is one such  method that is in widespread use for monitoring wildlife populations \citep{buckland2001introduction}. 
Among numerous applications \citep{buckland2016impact}, distance sampling has recently supported the monitoring of breeding bird populations across Europe \citep{rigal2023farmland} and marine mammal populations in the Eastern \citep{gilles2023scans} and Western \citep{roberts2016denmod} Atlantic Ocean. A distance sampling survey with line transects, which is the focus of this work, involves observers traversing straight lines and recording perpendicular distances to detected animals. 

The distance sampling methodology was developed in ecology over much of the past century, starting in earnest in the 1930s \citep[Section 1.10]{buckland2001introduction}. \citet{burnham1976mathematical} provided a key milestone in the development by giving a general mathematical formulation. The approach was further refined over the decades, and a thorough account of developments up to its publication are available in the textbook by \citet{buckland2001introduction}. A short time later, thought-provoking work by \citet{barry2001distance} appeared that challenged the status quo and suggested fundamental deficiencies in the distance sampling methodology.  
\citet{barry2001distance} demonstrated that the standard mathematical development of distance sampling has lacunae that can lead to estimation failures in even the simplest surveys (see their Table 1) and highlighted four ``underlying assumptions'' (see their Section 4.3) that generally do not hold. 
There was never a complete resolution to the matters raised by \citet{barry2001distance}\textemdash{see} \citet{welsh2002theory}, \citet{melville2001line}, \citet{fewster2005line} and \citet{melville2005line} for discussion and simulations. 
Each of the underlying assumptions of \citet{barry2001distance} are fair critiques of the existing theory, and in this work we provide a framework that either avoids or resolves them.

In statistical ecology there has recently been a renewed interest in identification of latent quantities, such as the unobservable animal abundance, in terms of the distribution of observable variables \citep[e.g.,][]{stoudt2023nonparametric,aleshin2024central}. The assumptions needed for an identification result reveal the extent to which the data are informative about the latent quantity of interest. 
Our first theorem establishes an identification result that underlies distance sampling. 

A major concern in ecological research is accommodating imperfect detection and detection heterogeneity, as generally some animals will not be detected, and detection probabilitity varies among animals  because of differences in animal proximity, behaviour and observer conditions. Many common approaches for abundance estimation, such as capture-recapture and distance sampling, are designed to accommodate that animals may not be detected despite being nearby and so available for detection. Capture-recapture, a very well known approach for abundance estimation, is sensitive to the probability of detection varying among animals and survey occasions \citep{link2003nonidentifiability,johndrow2019low,das2023doubly}. 
By contrast, it has been conjectured that distance sampling is not sensitive to detection heterogeneity. This conjecture was made by \citet[p. 45]{burnham1980estimation} and was given the name of \emph{pooling robustness} as it implies that detection probabilities can be pooled. 
The conjecture has been studied over the years but has never been fully resolved. \citet[Section 11.12]{buckland2004advanced} argued that Bayes rule, a key step in the identification proof for distance sampling, holds marginally when covariates that influence detection exist, thereby suggesting that heterogeneity can be ignored. This property of probability distributions is suggestive but is not itself a full proof. \citet{rexstad2023pooling} recently conducted simulations that strengthen the body of evidence in favor of distance sampling estimators having satisfactory performance in the presence of detection heterogeneity, but again this does not constitute a proof.  In this work, we demonstrate that the abundance identification result underlying distance sampling holds under arbitrary heterogeneity, thereby proving the pooling robustness conjecture and showing that distance sampling is ``immune'' to detection heterogeneity.

\subsection{Notation}

Consider $\N > 0$ animals at locations $\x_i \in \mathcal{E} \subseteq \R^2$, where $\mathcal{E}$ is some region of interest. Throughout, the animal locations $\x_1, \dots, \x_{\N}$ are considered fixed.

Consider \emph{transects} $\mS_j$ denoting line segments that are traversed during a search for animals, for $j=1, \dots, k$. Let $d$ be a distance and $d(e, \mS) := \inf_{s \in \mS} d(e, s)$ be the distance of location $e$ from a transect. 
Define $a(\mS_j; w) := \{e \in \mathcal{E} \, : \, d(e, \mS_j) \leq w \}$ as the covered area, or \emph{sampler}, associated with transect $j$. 
Each animal may or may not be detected on each transect: let $R_{ij}$ be an indicator for detection of animal $i$ on transect $j$. By construction, $R_{ij} = 0$ whenever $\x_i \not\in a(\mS_j; w)$. 
Define the distance $Y_{ij} := d(\x_i, \mS_j)$ of $\x_i$ from $\mS_j$. 
The observed data are $(\mS_j, \mY_j := \{ Y_{ij} \, : \, R_{ij}=1, i=1,\dots, \N \})$, that is, each transect $j=1, \dots k$ and the associated distances of the detected animals. Here and throughout the $\{\cdot\}$ notation denotes a multiset so permits co-occurrence of elements.
By construction, there is no information in the observed data about whether the same animal was detected multiple times across different transects.

\section{Identification}
\label{sec:identif}

In this section we identify the animal abundance $\N$ as a functional of the observed data. This crucial step enables estimation of $\N$, although we postpone development of estimation until the next section. The identification technique is non-standard relative to other missing data settings and carefully leverages all of the available information in the data and the study design. 
While this section presents interpretable and realistic assumptions, the proofs for this section are given under minimal but technical conditions contained in the Supplement. 

\subsection{Preliminaries}

On each transect, animals in $a(\mS_j; w)$ may or may not be detected. The detection functions 
\begin{equation*}
    g_{ij}(y) 
    := \P(R_{ij}=1 \mid Y_{ij}=y), 
\end{equation*}
for $j=1,\dots, k$ and $i=1,\dots, n$, describe the relevant components of this process. 
Let $\Pi_{ij}(y) :=  \P(Y_{ij} \leq y)$ for $y \geq 0$ be the distribution function of the distance $Y_{ij}$. 
We consider the following three assumptions. 
\begin{asp}[positivity]
\label{asp:pos}
    $\E\{ g_{ij}(Y_{ij})\}>0$ for
    all $i=1, \dots, \N$ and $j=1,\dots,k$. 
\end{asp}

\begin{asp}[representativeness]
\label{asp:repr}
     $\P \{ \x_i \in a(\mS_j; w) \}>0$ is known and does not depend on $i=1, \dots, \N$ for any $j=1, \dots, k$. 
\end{asp}

\begin{asp}[uniformity]
\label{asp:unif}
    $\Pi_{ij}$ is uniform over $[0,w]$ for all $i=1, \dots, \N$ and $j=1,\dots,k$.
\end{asp}
The positivity assumption ensures that each sampler is adequately surveyed, while the the representativeness and uniformity assumptions are on the transects and so the survey design. 
The first step towards identification of abundance is the following lemma, which at its core is a double inverse-probability weighted argument. 

\begin{lem}
\label{prop:ipw}
    If \Cref{asp:pos,asp:repr,asp:unif} 
    hold, then 
    \begin{equation*}
        \N 
        = \E \left[ \frac{1}{\frac{1}{kn} \sum_{j=1}^k \sum_{i=1}^n \P \{ \x \in a(\mS_j; w) \} \int_{(0,w]} g_{ij}(y) \,\d y / w} \frac{1}{k} \sum_{j=1}^k \# \mY_j \right],
    \end{equation*}
    where $x$ denotes an arbitrary animal location and $\#\mY$ denotes the cardinality of the set $\mY$. 
\end{lem}

The denominator of the prominent fraction in the lemma 
is an average over all transects and animals of the product of two probabilities: 
one is the coverage probability $\P \{ \x \in a(\mS_j; w) \}$, which corrects for sampling bias due to transect positioning, 
and the other is 
the average detection function $\frac{1}{w} \int_{(0,w]} g_{ij}(y) \,\d y$ over the covered area, which corrects for sampling bias due to detection. 
The conclusion of this inverse-probability weighting argument may seem routine, but it resolves one of the underlying assumptions highlighted by \citet{barry2001distance}.

\subsection{Identification}
\label{sec:identif:identif}

In many missing data or coarsening problems such as causal inference, the conclusion of the previous lemma 
would be an identified expression \citep{van2003unified}. However the same does not hold here since the aggregated detection function is not directly estimable. Indeed, the observed data is ``detection only'' in the sense that the distances are only observed when animals are detected. Although in other contexts estimation with such data is not possible without strong parametric assumptions \citep{hastie2013inference}, we show in this section that, in distance sampling, nonparametric identification is possible. 

The final key assumption is that animals on a transect $\mS_j$ are almost surely detected. 
\begin{asp}[perfect detection on the transect]
\label{asp:perf-detect}
    $g_{ij}(0) = 1$ for each $i=1, \dots, \N$ and $j=1, \dots, k$. 
\end{asp}
\noindent This assumption is typically valid in animal surveys, but can be violated for example during a visual shipboard survey of whales, which may be diving underwater while the boat passes overhead. 

Let $F_{ij}^*(y) := \P(Y_{ij} \leq y \mid R_{ij} = 1)$ be the distribution functions of detected distances and 
\begin{equation*}
    F^*(y):=\frac{\sum_{j=1}^k\sum_{i=1}^{\N}\P(R_{ij}=1)F_{ij}^*(y)}{\sum_{j=1}^k\sum_{i=1}^{\N}\P(R_{ij}=1)}
\end{equation*}
be a detection-probability weighted average of $F_{ij}^*$. 
The next result identifies the abundance $\N$ up to the derivative of the unknown function $F^*$ by applying Bayes rule. 

\begin{prop}
\label{thm:identif2}
    If \Cref{asp:pos,asp:repr,asp:unif,asp:perf-detect} hold,
    then 
    \begin{equation*}
        \N
        = \E \left\{ \frac{w}{\sum_{j=1}^k \P \{ \x \in a(\mS_j; w) \}} \f^*(0) \sum_{j=1}^k \# \mY_j \right\}, 
    \end{equation*}
    where $\f^*$ is the derivative of $F^*$. 
\end{prop}

The proportion of ``close'' (i.e., with $y \leq u$) detections is 
\begin{equation}
    F_{\no}(u)
    := \frac{1}{\sum_{j=1}^k \E \left( \# \mY_j \right)} \sum_{j=1}^k \E \left\{ \sum_{y \in \mY_j} I (y \leq u) \right\}, \label{eq:f}
\end{equation}
and we show that this identifies $F^*(u)$. 
\begin{lem}
\label{lem:f}
    If \Cref{asp:pos} 
    holds, then $F(u) = F^*(u)$ for $u \in [0,w]$. 
\end{lem}

\begin{rem}
    The function $F$ satisfies the criterion of \emph{pooling robustness} in distance sampling since $F$ is robust to pooling animals with different detection functions together. We must introduce some notation to state this in more detail. Suppose that the $\N$ animals may be partitioned into $r$ strata and that the observed data for each transect are detected distances in each stratum $\mY^{(1)}, \dots, \mY^{(r)}$ rather than in aggregate. Denote $F^{(s)}$ as the version of $F$ calculated only among animals in stratum $s$, i.e., with data $\mY^{(s)}_1, \dots, \mY^{(s)}_k$. We may readily establish that for $y \in [0,w]$, 
    \begin{equation}
        F(y)
        = \sum_{s=1}^r \frac{\sum_{j=1}^k \E(\#\mY^{(s)}_j)}{\sum_{s'=1}^r \sum_{j=1}^k \E(\#\mY^{(s')}_j)} F^{(s)}(y), \label{eq:pr}
    \end{equation}
    which states that $F$ is invariant to pooling: after calculating $F^{(s)}$ in each stratum $s=1, \dots, r$, the aggregate $F$ may be calculated by an average weighted by detection counts. 
    When all expressions in \Cref{eq:pr} are replaced with their sample analogs, the resulting equation is the pooling robustness criterion \citep[p. 45]{burnham1980estimation}. The criterion was motivated by considerations that abundance estimates within each stratum should aggregate to the overall abundance estimate. 
\end{rem}

\begin{thm}
\label{thm:identif}
    If \Cref{asp:repr,asp:unif,asp:pos,asp:perf-detect} hold, 
    then 
    \begin{equation}
        \N 
        = \E \left[ \frac{w}{\sum_{j=1}^k \P \{ \x \in a(\mS_j; w) \}} \f_{\no}(0) \sum_{j=1}^k \# \mY_j \right], \label{eq:identif}
    \end{equation}
    where $\f_{\no}$ is the derivative of $F_{\no}$ defined in \Cref{eq:f} 
\end{thm}

The result is striking as it makes no assumptions about heterogeneity in detection. 
In the case of detection homogeneity so that $g_{ij}$ does not depend on $i,j$, some version of this result is widely believed in the distance sampling community to be true, however as far we are aware this is the first result to rigorously establish it.

\section{A transect distribution example}
\label{sec:example}

In this section, we consider a simple bounded environment to illustrate the previous material and determine transect distributions that satisfy the design assumptions. As far as we are aware, this section contains the first example of a transect distribution that provides identification of animal abundance in distance sampling. 
To simplify result statements, we consider only $k=1$ transect although the conclusions remain unchanged when $k>1$. 

Let $\mathcal{E} \subset \R^2$ be a bounded set. Fix an orthonormal basis $(x,y)$, and write the coordinates of a point $e\in\R^2$ in this basis as $e = (e_x,e_y)$. Define
\begin{equation*}
    x_{\lo} := \inf\{e_x : e\in \mathcal{E}\}, 
    \qquad
    x_{\hi} := \sup\{e_x : e\in \mathcal{E}\},
\end{equation*}
and assume the distance $d(\cdot,\cdot)$ is the usual Euclidean distance.
Consider the random transect
\begin{equation*}
    \mS = \{ (S_x, s_y) \, : \, y_{\lo}(S_x) \leq s_y \leq y_{\hi}(S_x) \},
\end{equation*}
where the normal offset $S_x$ has some distribution and the vertical extents of the transects are 
\begin{equation}
\label{eq:yloyhi-def}
    y_{\lo}(s) := \inf\{e_y : e\in\mathcal{E},\ |e_x - s|\leq w\},
    \qquad
    y_{\hi}(s) := \sup\{e_y : e\in\mathcal{E},\ |e_x - s|\leq w\}.
\end{equation}

To connect with the previous assumptions, note that for any animal location $e=(e_x,e_y)\in \mathcal{E}$ and any $y\in[0,w]$,
\begin{equation}
\label{eq:dist-indicator}
    \{ d(e,\mS) \le y \}
    \;=\;
    \{ |e_x - S_x| \le y \}.
\end{equation}
Indeed, if $d(e,\mS)\le y$ then necessarily $|e_x-S_x|\le y$ by the Euclidean lower bound on distance. Conversely, if $|e_x-S_x|\le y\le w$ then $e\in\{u\in\mathcal{E}: |u_x-S_x|\le w\}$, so $e_y\in[y_{\lo}(S_x),y_{\hi}(S_x)]$ by definition \eqref{eq:yloyhi-def}; hence $(S_x,e_y)\in\mS$ and $d(e,\mS)=|e_x-S_x|\le y$.
\begin{prop}
\label{prop:no-vert}
    If the support of $S_x$ contains $(x_{\lo},x_{\hi})$ and $\P ( |e_x - S_x| \leq y )$ does not depend on $e_x \in (x_{\lo}+y,\,x_{\hi}-y)$ for all $y\in[0,w]$, then $S_x$ is uniform over $(x_{\lo},x_{\hi})$. 
\end{prop}

\begin{proof}[Proof of \Cref{prop:no-vert}]
    It is clear that $S_x$ must have a Lebesgue density, say $h$.
    
    Consider any two points $u_1, u_2 \in [x_{\lo}+y,\,x_{\hi}-y]$ such that $|u_1 - u_2| \leq 2w$. Then the expression
    \begin{equation*}
        \P \left\{ \left|\frac{u_1+u_2}{2} - S_x \right| \leq \frac{|u_1 - u_2|}{2} \right\}
        = \int_{\left[ e_x - \frac{|u_1 - u_2|}{2},\, e_x + \frac{|u_1 - u_2|}{2} \right]} \, h(u)\,\d u
    \end{equation*}
    does not depend on $e_x=\frac{u_1+u_2}{2}$, by assumption, so its derivative with respect to $e_x$ vanishes. That is, 
    \begin{equation}
        h(u_2) - h(u_1)
        = 0,
    \end{equation}
    showing that the density of $S_x$ is equal at $u_1$ and $u_2$. 
    
    It is clear that for any two points $u_1,u_2 \in (x_{\lo},x_{\hi})$, there exists a sequence of points $u_3, \dots, u_m$, for some integer $m$, such that $u_1 < u_3 < \cdots < u_m < u_2$ and each neighboring point (e.g. $u_1$ and $u_3$) satisfy the premise of the above construction. Thus $h(u_1) = h(u_2)$, and $h$ is uniform over $(x_{\lo},x_{\hi})$. 
\end{proof}

\noindent A corollary is that the uniformity of $S_x$ over $(x_{\lo},x_{\hi})$, which is suggestively connected to the uniformity assumption, is a necessary condition for the representativeness assumption to hold. 

The following establishes a transect distribution that attains the design assumptions. 

\begin{thm}
\label{prop:luigi2}
    The distribution $S_x \sim \mathrm{Uniform}(x_{\lo}-w,\,x_{\hi}+w)$ satisfies the uniformity and representativeness assumptions.
\end{thm}

\begin{proof}[Proof of \Cref{prop:luigi2}]
    For any $e_x \in [x_{\lo},x_{\hi}]$, the coverage probability
    \begin{equation*}
        \P ( |e_x - S_x| \leq y ) = \frac{2y}{(x_{\hi}-x_{\lo})+2w}
        \qquad \text{for any $y \in [0,w]$},
    \end{equation*}
    which does not depend on $e_x$. This shows \Cref{asp:unif} holds by considering $e_x = x_1, \dots, x_{\N}$, and the equivalence \eqref{eq:dist-indicator} yields the corresponding representativeness statement up to distance $w$.
\end{proof}

Edge effects are a well-known issue in distance sampling \citep[see][p. 200]{buckland2004advanced}. The above transect distribution employs \emph{plus sampling} to allow $\mS$ to be outside but near the edges of $\mathcal{E}$ \citep{buckland2006point2}.

\section{Estimation}
\label{sec:est}

The identification result in \Cref{sec:identif:identif} shows that the estimand 
\begin{equation*}
    \psi
    := \E \left[ \frac{w}{\sum_{j=1}^k \P \{ \x \in a(\mS_j; w) \}} \f_{\no}(0) \sum_{j=1}^k \# \mY_j \right] 
\end{equation*}
characterises the abundance $n$ in terms of the distribution of the observed data $\mY_1, \dots, \mY_k$. 
In this section we study estimation and inference for $\psi$. 

\subsection{Efficiency bound}

A key component of $\psi$ is the density $\f(0)$ evaluated at $0$. Considering the definition of $F$ in \Cref{eq:f}, only detected distances in a neighborhood of $0$ contribute to the value of $f(0)$ in the absence of structural assumptions. Formally, this makes $f(0)$ not pathwise differentiable as a functional of the observed data distribution \citep[Section 3.3]{bickel1993efficient}. Standard theory for the estimation and the efficiency bounds of statistical functionals consequently does not apply \citep[cf.][]{kennedy2022semiparametric}. Additionally, the value $0$ is on the boundary of the domain of $f$, which further complicates estimation. 

In the distance sampling community, parametric approximations to $f$ are commonly employed in part due to the relatively small sample sizes commonly experienced in practice. 
This perspective is based on decades of empirical work, so we use it to guide our estimation strategy while maintaining a nonparametric model. 
Consider a smooth parametric curve $F_{\no}(\cdot; \theta)$ containing an element that approximates $F_{\no}$. 
For $f(y; \theta) = \frac{\d}{\d y} F(y; \theta)$, define $\theta_0$ so that 
\begin{equation}
    \theta_0
    := \arg\max_{\theta} \E \left\{ \sum_{j=1}^k \sum_{Y \in \mY_j} \log \f_{\no}(Y; \theta) \right\}, \label{eq:beta0-defn}
\end{equation}
which we assume exists and is unique. 
\noindent The derivative $f(\cdot; \theta_0)$ is a projection of $f$ onto the parametric curve. 

Define the modified estimand 
\begin{equation*}
    \tilde\psi 
    := \frac{w}{\sum_{j=1}^k \P \{ \x \in a(\mS_j; w) \}} \f_{\no}(0; \theta_0) \sum_{j=1}^k \E \left( \# \mY_j \right)
    = \frac{f(0; \theta_0)}{f(0)} \psi 
\end{equation*}
which contains the projection-based term $f(0; \theta_0)$ rather than $f(0)$. In general $\psi \neq \tilde\psi$, however when $\tilde\psi$ is estimated this approximation error may be small relative to any estimation error. In other settings, this strategy of replacing a non-pathwise differentiable expression (here the evaluated density $f(0)$) with a projection has previously been employed \citep[see, e.g.,][]{neugebauer2007nonparametric,van2023adaptive}. 
In the remainder of this section we study the operating characteristics of estimators for $\tilde\psi$ as a proxy for $\psi$.

We consider the asymptotic behaviour of estimators $\hat\psi_k$ for $\tilde\psi$ as the number of transects $k \to \infty$. Since $k$ corresponds to the sampling effort, this is analogous to standard asymptotics which consider the ``sample size'' diverging. 
The following theorem characterizes the efficiency bound for $\tilde\psi$ and characterises an estimator that attains the bound. 
Throughout we consider the model at hand: the distribution of the transect $\mS_j$ is known while the conditional distribution of the detected distances $\mY_j$ given the transect $\mS_j$ is nonparametric. 

\begin{thm}
\label{thm:est-working}
    Under independent sampling, the class of influence functions for estimating $\tilde{\psi}(\PP)$ is $\varphi(\mY; \PP) - [ h(\mS) - \E\{h(\mS)\} ]$, where $h$ is an arbitrary finite-variance function and 
    \begin{align*}
        \varphi(\mY; \PP)
        & := \frac{w}{\P \{ \x \in a(\mS; w) \}} \biggl[ \f_{\no}(0; \theta_0) \Bigl\{ \# \mY - \E \left( \# \mY \right) \Bigr\} - \\
        & \hspace{40mm} \E \left( \# \mY \right) \biggl\{
         \dot{\f}(0; \theta_0)^{\top} V(\theta_0)^{-1} \sum_{Y \in \mY} \frac{\dot{\f}(Y; \theta_0)}{\f(Y; \theta_0)} \biggr\} \biggr]. 
    \end{align*}
    Here $V(\theta) := \frac{\d^2}{\d \theta \d \theta^{\top}} \E \left\{ \sum_{Y \in \mY} \log\f(Y; \theta) \right\}$ and $\dot{\f}(\cdot; \theta) := \frac{\partial}{\partial\theta} \f(\cdot; \theta)$. 
    Further, the efficient influence function of $\tilde\psi(\PP)$ has $h_{\text{eff}}(\mS) := \E\{\varphi(\mY; \PP) \mid \mS\}$. 
\end{thm}

\subsection{Estimation}

The conventional estimator for the detection parameter $\theta_0$ is the $M$-estimator given by 
\begin{equation*}
    \hat\theta_k
    := \arg\max_{\theta} \left\{ \sum_{j=1}^k \sum_{Y \in \mY_j} \log \f_{\no}(Y; \theta) \right\},
\end{equation*}
which parallels the definition of $\theta_0$ in \Cref{eq:beta0-defn}. 
The standard estimator for $\tilde\psi$ is the plug-in estimator 
\begin{equation*}
    \hat\psi_k^{\text{std}}
    := \frac{w}{\sum_{j=1}^k \lambda \{ a(\mS_j; w) \} / \lambda ( \mathcal{E} )} \f_{\no}(0; \hat\theta_k) \sum_{j=1}^k \# \mY_j,
\end{equation*}
where the relative area in the denominator has expectation 
\begin{equation*}
    \E \left[ \frac{\lambda \{ a(\mS_j; w) \}}{\lambda ( \mathcal{E} )} \right]
    = \frac{1}{\lambda ( \mathcal{E} )} \E \left[ \int_{\mathcal{E}} I \Bigl\{ x \in a(\mS_j; w) \Bigr\} \,\d x \right]
    = \P \Bigl\{ x \in a(\mS_j; w) \Bigr\}
\end{equation*}
under the assumption that coverage probability (on the RHS) does not depend on the animal location $x$, which is slightly 
stronger than \Cref{asp:repr}. 

In the next result we compare the influence functions of the standard estimator and the natural plugin estimator.
\begin{prop}
\label{prop:est-review}
    Under independent sampling and standard regularity conditions, 
    the influence function of the standard estimator $\hat\psi_k^{\text{std}}$ has augmentation term 
    $h(\mS) = \frac{\tilde{\psi}}{\P \{ \x \in a(\mS; w) \}} \frac{\lambda \{ a(\mS; w) \}}{\lambda ( \mathcal{E} )}$,
    and the influence function of the plugin estimator 
    \begin{equation*}
        \frac{w}{\sum_{j=1}^k \P \{ \x \in a(\mS_j; w) \}} \f_{\no}(0; \hat\theta_k) \sum_{j=1}^k \# \mY_j 
    \end{equation*}
    has augmentation term $0$. 
\end{prop}

\noindent Neither influence function uniformly has lower variance than the other, and both are generally distinct from the efficient influence 
function.

Based on the form of the efficient influence function, we propose the augmented estimator
\begin{equation*}
    \hat\psi_k
    := \frac{w}{\sum_{j=1}^k \P \{ \x \in a(\mS_j; w) \}} \f_{\no}(0; \hat\theta_k) \sum_{j=1}^k \# \mY_j 
        - \frac{1}{k} \sum_{j=1}^k \widehat{m}_k (\mS_j),
\end{equation*}
where $\widehat{m}_k(\cdot)$ estimates a linear approximation to the efficient augmentation term $h_{\text{eff}}(\mS) := \E\{\varphi(\mY; \PP) \mid \mS\}$. 
Let $\widehat\varphi_k(\cdot)$ be the natural plugin estimator of the influence function $\varphi(\cdot)$, and let $t(\mS)$ denote a
finite dimensional summary of the transect (e.g. its sampler area, its position in the environment, and transformations). 
Since the transect distribution is known, we center the summary so that $\E\{t(\mS)\}=0$. In particular, we put 
$\widehat{m}_k(\mS) := \widehat\beta_k^{\top}t(\mS)$ where the coefficient is 
\begin{equation*}
    \widehat\beta_k
    := \left\{ \sum_{j=1}^k t(\mS_j)t(\mS_j)^{\top} \right\}^{-1} \sum_{j=1}^k t(\mS_j)\widehat\varphi_k(\mY_j),
\end{equation*}
the linear regression of the estimated influence function on the centered transect summary without an intercept. 
Denote the corresponding population projection by $m_t(\mS) := \beta_t^{\top}t(\mS)$ where the coefficient is 
$\beta_t := \arg\min_{\beta} \E\left[\left\{\varphi(\mY;\PP)-\beta^{\top}t(\mS)\right\}^2\right]$. 

\begin{thm}
\label{thm:est}
    Under independent sampling and standard regularity conditions, the estimator $\hat\psi_k$ satisfies the asymptotic expansion 
    \begin{equation*}
        \hat\psi_k(\mS_1, \mY_1; \dots; \mS_k, \mY_k) - \tilde\psi(\PP) 
        = \frac{1}{k} \sum_{j=1}^k \left\{ \varphi(\mY_j; \PP) - m_t(\mS_j) \right\} + o_{\PP}(k^{-1/2}). 
    \end{equation*}
\end{thm}

\noindent The proposed estimator is not generally efficient but will be when the efficient augmentation term $h_{\text{eff}}(\mS) := \E\{\varphi(\mY; \PP) \mid \mS\}$
is linear in the centered summary $t(\mS)$. In practice, the expectation may be high dimensional which is not feasible to estimate with the
dozen or two transects in typical surveys. 
Critically, the estimator $\hat\psi_k$ is $\sqrt{k}$-consistent regardless of whether augmentation regression is correctly
specified.

The asymptotic linearity result makes possible inference for $\tilde\psi$. Denote $\hat\varphi_k$ as the influence function $\varphi$ with all terms replaced with their sample analogs so that $\hat\varphi_k$ is consistent for $\varphi$, and put $\hat\varphi_k^{\mathrm{aug}}(\mS,\mY):=\hat\varphi_k(\mY)-\widehat m_k(\mS)$. Then, by the central limit theorem,
\begin{equation*}
    \hat\psi_k \approx \mathcal{N} \left[ \tilde\psi, \frac{1}{k^2} \sum_{j=1}^k \left\{ \hat\varphi_k^{\mathrm{aug}}(\mS_j,\mY_j) \right\}^2 \right], 
\end{equation*}
where here ``$\approx$'' denotes a distribution approximation. 
When $\tilde\psi > 0$, the delta method with the map $u \mapsto \log u$ gives
\begin{equation*}
    \log \hat\psi_k \approx \mathcal{N} \left[ \log \tilde\psi, \frac{1}{k^2} \sum_{j=1}^k \left\{ \frac{\hat\varphi_k^{\mathrm{aug}}(\mS_j,\mY_j)}{\hat\psi_k} \right\}^2 \right], 
\end{equation*}
so that a log Wald confidence interval is obtained by exponentiating the corresponding symmetric confidence interval on the log scale. This has the practical advantages that the resulting interval is always positive and is better aligned with the right-skewed sampling behaviour often seen for abundance estimators in small samples. 

\section{Discussion} 
\label{sec:disc}

In this work we have developed a foundation for the distance sampling methodology. Readers unfamiliar with distance sampling should find a thorough and satisfactory account of the statistical theory for a common statistical survey method in ecology. 
Readers familiar with distance sampling should find four primary differences relative to the standard presentation. 
First, practical assumptions are provided which leads to the resolution of the first three underlying assumptions of \citet{barry2001distance}. 
Second, an alternative estimator for the variance of the abundance estimator $\hat\psi_k$ is provided that takes into account the dependence between estimator components 
and thereby resolves the last underlying assumption of \citet{barry2001distance}.
Third, the identification result in \Cref{sec:identif:identif} contains the coverage probability $\P \{ \x \in a(\mS_j; w) \}$, which averages over the 
distribution of $\mS_j$, whereas the standard approach instead contains a related empirical term proportional to the area $\lambda \{ a(\mS_j; w) \}$ of the sampler; 
the two expressions should be similar when $k$ is not small. 
Finally, the results show that identification and consistency of the distance sampling estimator is unaffected by detection heterogeneity, resolving a decades-long conjecture \citep{burnham1980estimation,rexstad2023pooling}. 

Overall, the work makes clear which assumptions underlie distance sampling and consequently which assumptions do not; 
for example, the ecology literature often erroneously reports that each detection function $g_{ij}$ decreasing is a necessary assumption \citep[see, e.g.,][Section 3.2.5]{king2014statistical}. 
We have also proposed a nearly efficient abundance estimator. 
Distance sampling is a rich area, and we have only touched on the core of it. The work here has many implications, including on 
Horvitz-Thompson estimation \citep[see, e.g.,][Section 2.3]{buckland2001introduction},
variance estimation \citep{fewster2009estimating}, 
mark-recapture distance sampling \citep{borchers1998mark},
covariate-informed detection \citep{marques2003incorporating}, and
non-linear transects.

\bibliographystyle{apalike}
\bibliography{refs}

\appendix
\makeatletter
\renewcommand\thesection{S\@arabic\c@section}
\renewcommand{\theequation}{S\arabic{equation}}
\renewcommand\theasp{S\arabic{asp}}
\renewcommand\thelem{S\arabic{lem}}
\renewcommand\thethm{S\arabic{thm}}
\renewcommand\theprop{S\arabic{prop}}
\renewcommand\theHasp{S.\arabic{asp}}
\renewcommand\theHlem{S.\arabic{lem}}
\renewcommand\theHthm{S.\arabic{thm}}
\renewcommand\theHprop{S.\arabic{prop}}
\makeatother
\section{Proofs for \texorpdfstring{\Cref{sec:identif}}{identification section}}
\label{sec:supp:identif-proofs}

\subsection{Results under weaker assumptions}

Throughout the proofs for this section we consider the following assumptions alternative to \Cref{asp:repr,asp:unif}. Define $\Pi_{\bullet\bullet} := \frac{1}{kn} \sum_{j=1}^k \sum_{i=1}^n \Pi_{ij}$. Denote $\d\Pi$ as the density of $\Pi$ with respect to an implicit/implied dominating measure, and in general let $\d$ be a density operator that gives the density of its argument. 

\begin{asp}
\label{asp:repr2}
    There exists some $i=1,\dots,n$ and $j=1,\dots,k$ such that $\d\Pi_{ij}(0)>0$. 
\end{asp}

\begin{asp}
\label{asp:unif2}
    $\d \Pi_{\bullet\bullet}(0)$ is known. 
\end{asp}

The next lemma provides a connection between \Cref{asp:unif2,asp:repr2} and \Cref{asp:repr,asp:unif}. 
\setcounter{lem}{-1}
\begin{lem}
\label{asp:app}
    Consider a randomised transect such that $\P \{ \x_i \in a(\mS_j; w) \}$ is known for each $i=1,\dots,n$. 
    If \Cref{asp:repr,asp:unif} hold, then \Cref{asp:unif2,asp:repr2} hold. 
\end{lem}

\begin{proof}
    Throughout we will use the identity 
    \begin{equation}
        \Pi_{ij}(y)
        = \P(Y_{ij} \leq y \mid Y_{ij} \leq w) \P(Y_{ij} \leq w)
        = \frac{y}{w} \P \{ \x \in a(\mS_j; w) \}, \label{eq:pi-formula}
    \end{equation}
    where the last equality applies the premises. 
    \Cref{asp:repr2} holds because $\d \Pi_{ij}(0) = \P \{ \x \in a(\mS_j; w) \}/w > 0$ for each $i,j$. 
    \Cref{asp:unif2} holds because $\d \Pi_{ij}(0) = \P \{ \x \in a(\mS_j; w) \}/w$ is known for each $i,j$. 
\end{proof}

We now start the theoretical development. Denote $G_{ij}(y) := \int_{(0,y]} g_{ij}(u) \,\d \Pi_{ij}(u)$, $G_{\bullet\bullet} := \frac{1}{kn} \sum_{j=1}^k \sum_{i=1}^n G_{ij}$, and 
\begin{equation}
\label{eq:Fstardefn}
    F^*(y) := \sum_{j=1}^k \sum_{i=1}^n G_{ij}(w) F_{ij}(y) \Big/ \sum_{j=1}^k \sum_{i=1}^n G_{ij}(w); 
\end{equation}  
under \Cref{asp:unif}, this definition reduces to that considered in the main text. 

\begin{lem}[\Cref{prop:ipw}]
\label{s-prop:ipw}
    If \Cref{asp:pos} 
    holds, then 
    \begin{equation*}
        \N 
        = \E \left\{ \frac{\sum_{j=1}^k \#\mY_j}{k G_{\bullet \bullet}(w)} \right\}, 
    \end{equation*}
    where $x$ denotes an arbitrary animal location and $\#\mY$ denotes the cardinality of the set $\mY$. 
\end{lem}

\begin{proof}[Proof of \Cref{s-prop:ipw}]
    The marginal detection probability is 
    \begin{align*}
        \E ( R_{ij}) 
        & = \E \{ \E (R_{ij} \mid Y_{ij} ) \} 
        = \int_{(0,w]} \E (R_{ij} \mid Y_{ij}=y) \,\d \P(Y_{ij} \leq y) \\
        & = \int_{(0,w]} g_{ij}(y) \,\d \Pi_{ij}(y). 
    \end{align*}
    Summing this expression and applying the identity $\sum_{j=1}^k \#\mY_j = \sum_{j=1}^k \sum_{i=1}^n R_{ij}$, we get
    \begin{equation*}
        \E \left( \sum_{j=1}^k \#\mY_j \right)
        = \sum_{j=1}^k \sum_{i=1}^n G_{ij}(w),
    \end{equation*}
    where we recall that we denote $G_{ij}(y) := \int_{(0,y]} g_{ij}(u) \,\d \Pi_{ij}(u)$. Now, with the definition $G_{\bullet\bullet} := \frac{1}{kn} \sum_{j=1}^k \sum_{i=1}^n G_{ij}$, we get
    \begin{equation*}
        n
        = \E \left\{ \frac{\sum_{j=1}^k \#\mY_j}{k G_{\bullet \bullet}(w)} \right\},
    \end{equation*}
    since $G_{\bullet \bullet}(w) = \frac{1}{kn} \sum_{j=1}^k \sum_{i=1}^n \P(R_{ij}=1) > 0$ under \Cref{asp:pos}. 
\end{proof}

\begin{lem}
\label{prop:p-identif}
    If \Cref{asp:perf-detect,asp:repr2} hold, then for $y \in [0,w]$ 
    \begin{equation*}
        \frac{\d G_{\bullet\bullet}(y)}{\d \Pi_{\bullet\bullet}(0)}
        = \frac{\f^*(y)}{\f^*(0)}, 
    \end{equation*}
    where $\f^*$ is the derivative of $F^*$. 
\end{lem}

\begin{proof}[Proof of \Cref{prop:p-identif}]
    By Bayes rule we readily see that 
    \begin{equation*}
        \,\d \P(Y_{ij} \leq y \mid R_{ij} = 1) 
        = \frac{\P(R_{ij} = 1 \mid Y_{ij} = y)\,\d \P(Y_{ij} \leq y)}{\int_{(0,w]} \P(R_{ij} = 1 \mid Y_{ij} = u) \,\d \P(Y_{ij} \leq u)} 
    \end{equation*}
    where the integral in the denominator is over $(0,w]$ since we have defined $R_{ij}$ such that $\P(R_{ij} = 1 \mid Y_{ij} = u) = 0$ for $u > w$. 
    Expressing this display in terms of the previously defined notation, we get 
    \begin{equation}
        f_{ij}(y) 
        = \frac{g_{ij}(y)\,\d \Pi_{ij}(y)}{G_{ij}(w)}. \label{eq:bayes} 
    \end{equation}
    Taking $y=0$ in \cref{eq:bayes} and applying \Cref{asp:perf-detect}, we get 
    \begin{equation}
        f_{ij}(0) 
        = \frac{\d \Pi_{ij}(0)}{G_{ij}(w)}. \label{eq:bayes0} 
    \end{equation}
    Now, combining \cref{eq:bayes,eq:bayes0} and summing, we get 
    \begin{align*}
        \frac{\d G_{\bullet\bullet}(y)}{\d \Pi_{\bullet\bullet}(0)} 
        & = \frac{\sum_{j=1}^k \sum_{i=1}^n G_{ij}(w) f_{ij}(y)}{\sum_{j=1}^k \sum_{i=1}^n G_{ij}(w) f_{ij}(0)} \\ 
        & = \frac{\sum_{j=1}^k \sum_{i=1}^n G_{ij}(w) f_{ij}(y) \Big/ \sum_{j=1}^k \sum_{i=1}^n G_{ij}(w)}{\sum_{j=1}^k \sum_{i=1}^n G_{ij}(w) f_{ij}(0) \Big/ \sum_{j=1}^k \sum_{i=1}^n G_{ij}(w)} \\
        & = \frac{f^*(y)}{f^*(0)}, 
    \end{align*}
    where the combination applied \Cref{asp:repr2} to ensure $\d \Pi_{\bullet\bullet}(0)>0$. 
\end{proof}

\begin{prop}[\Cref{thm:identif2}]
\label{s-thm:identif2}
    If \Cref{asp:pos,asp:repr2,asp:perf-detect} hold,
    then 
    \begin{equation*}
        \N
        = \E \left\{ \frac{f^*(0) \sum_{j=1}^k \#\mY_j}{k \,\d \Pi_{\bullet\bullet}(0)} \right\}, 
    \end{equation*}
    where $\f^*$ is the derivative of $F^*$ defined in \Cref{eq:Fstardefn}. 
\end{prop}

\begin{proof}[Proof of \Cref{s-thm:identif2}]    
    Applying \Cref{s-prop:ipw}, we get 
    \begin{equation*}
        n
        = \E \left\{ \frac{\sum_{j=1}^k \#\mY_j}{k G_{\bullet \bullet}(w)} \right\},
    \end{equation*}
    Applying \Cref{prop:p-identif} and integrating, we get 
    \begin{equation*}
        \int_{(0,w]} \frac{\d G_{\bullet\bullet}(y)}{\d \Pi_{\bullet\bullet}(0)} 
        = \int_{(0,w]} \frac{f^*(y)}{f^*(0)} \,\d y 
        = \frac{1}{f^*(0)}, 
    \end{equation*}
    since $\int_{(0,w]} f_{ij}(u) \,\d u = 1$ as all detected animals are within distance $w$. 
    Plugging this into the previous display completes the argument. 
\end{proof}

The following lemma supports the proof of \Cref{s-lem:f}. 

\begin{lem}
\label{lem:flike-identif}
    If $\P(R_{ij}=1)>0$ for some $i,j$, then 
    for any $u \in [0,w]$, 
    \begin{align*}
        \frac{\sum_{j=1}^k \sum_{i=1}^{\N} \P \left( R_{ij}=1 \right) F_{ij}^*(u)}{\sum_{j=1}^k \sum_{i=1}^{\N} \P \left( R_{ij}=1 \right)}
        = \frac{1}{\sum_{j=1}^k \E \left( \# \mY_j \right)} \sum_{j=1}^k \E \left\{ \sum_{y \in \mY_j} I (y \leq u) \right\}. 
    \end{align*}
\end{lem}

\begin{proof}[Proof of \Cref{lem:flike-identif}]
    We separately show the numerators and denominators are equal. 
    The denominator satisfies 
    \begin{align*}
        \sum_{j=1}^k \left\{ \sum_{i=1}^{\N} \P \left( R_{ij}=1 \right)\right\} 
        = \sum_{j=1}^k \left\{ \sum_{i=1}^{\N} \E \left( R_{ij} \right) \right\} 
        = \sum_{j=1}^k \E \left( \# \mY_j \right), 
    \end{align*}
    while the numerator satisfies 
    \begin{align*}
        \sum_{j=1}^k \E \left\{ \sum_{y \in \mY_j} I (y \leq u) \right\}  
        & = \sum_{j=1}^k \E \left\{ \sum_{i=1}^{\N} R_{ij} I(Y_{ij} \leq u) \right\} \\
        & \hspace{-15mm} = \sum_{j=1}^k \sum_{i=1}^{\N} \E \left\{ R_{ij} I(Y_{ij} \leq u) \right\} \\
        & \hspace{-15mm} = \sum_{j=1}^k \sum_{i=1}^{\N} \P \left( Y_{ij} \leq u \middle| R_{ij}=1 \right) \P ( R_{ij}=1 ) \\ 
        & \hspace{-15mm} = \sum_{j=1}^k \sum_{i=1}^{\N} F_{ij}^*(u) \P ( R_{ij}=1 ). 
    \end{align*}
    The result follows as the division is well-defined by the assumption. 
\end{proof}

\begin{lem}[\Cref{lem:f}]
\label{s-lem:f}
    If \Cref{asp:pos} 
    holds, then $F(u) = F^*(u)$ for $u \in [0,w]$. 
\end{lem}

\begin{proof}[Proof of \Cref{s-lem:f}]
    Under \Cref{asp:pos}, we get  $\P(R_{ij}=1)>0$ for all $i,j$. 
    Since $G_{ij}(w) = \P(R_{ij}=1)$, we get  
    \begin{align*}
        \frac{\sum_{j=1}^k \sum_{i=1}^n G_{ij}(w) F_{ij}^*(y)}{\sum_{j=1}^k \sum_{i=1}^n G_{ij}(w)} 
        = \frac{\sum_{j=1}^k \sum_{i=1}^n \P(R_{ij}=1) F_{ij}^*(y)}{\sum_{j=1}^k \sum_{i=1}^n \P(R_{ij}=1)}. 
    \end{align*}
    By applying \Cref{lem:flike-identif}, we get for $u \in (0,w]$ 
    \begin{equation*}
        \frac{\sum_{j=1}^k \sum_{i=1}^{\N} \P \left( R_{ij}=1 \right) F_{ij}^*(u)}{\sum_{j=1}^k \sum_{i=1}^{\N} \P \left( R_{ij}=1 \right)}
        = F(u),
    \end{equation*}
    as desired. 
\end{proof}

\begin{thm}[\Cref{thm:identif}]
\label{s-thm:identif}
    If \Cref{asp:pos,asp:repr2,asp:unif2,asp:perf-detect} hold, 
    then 
    \begin{equation}
        \N 
        = \E \left\{ \frac{f(0) \sum_{j=1}^k \#\mY_j}{k \,\d \Pi_{\bullet\bullet}(0)} \right\}, \label{s-eq:identif}
    \end{equation}
    is identified, 
    where $\f_{\no}$ is the derivative of $F_{\no}$ defined in \Cref{eq:f}. 
\end{thm}

\begin{proof}[Proof of \Cref{s-thm:identif}]
    It follows directly from \Cref{s-thm:identif2} by plugging in the equivalence of $F^*$ and $F_{\no}$ from \Cref{s-lem:f} and then applying \Cref{asp:unif2} for identification. 
\end{proof}

\subsection{Results as presented in the main document}

In this section we specialise the results in the previous section to the case when \Cref{asp:repr,asp:unif} hold. We repeatedly apply the following two simplifications.
\begin{lem}
    Let $y \in (0,w]$. 
    If \Cref{asp:repr,asp:unif} hold, then 
    \begin{equation}
        G_{\bullet \bullet}(y)
        = \frac{1}{kn} \sum_{j=1}^k \sum_{i=1}^n \frac{\P \{ \x \in a(\mS_j; w) \}}{w} \int_{(0,y]} g_{ij}(u) \,\d u. \label{eq:gdd-simp}
    \end{equation}
    If \Cref{asp:unif} holds, then  
    \begin{equation}
        \d\Pi_{ij}(y)
        = \frac{1}{w} \P \{ \x_i \in a(\mS_j; w) \}. \label{eq:pi-simp}
    \end{equation}
\end{lem}
\begin{proof}
    First, 
    \begin{align*}
        G_{\bullet \bullet}(y)
        & = \frac{1}{kn} \sum_{j=1}^k \sum_{i=1}^n \int_{(0,y]} g_{ij}(u) \,\d \Pi_{ij}(u) \nonumber \\ 
        & \hspace{-5mm} = \frac{1}{kn} \sum_{j=1}^k \sum_{i=1}^n \frac{\P \{ \x \in a(\mS_j; w) \}}{w} \int_{(0,y]} g_{ij}(u) \,\d u.
    \end{align*}
    Second, 
    \begin{equation*}
        \Pi_{ij}(y)
        = \P(Y_{ij} \leq y \mid Y_{ij} \leq w) \P(Y_{ij} \leq w)
        = \frac{y}{w} \P \{ \x_i \in a(\mS_j; w) \}. \qedhere
    \end{equation*}
\end{proof}

Now we present the proofs of the main results. 

\begin{proof}[Proof of \Cref{prop:ipw}]
    Under \Cref{asp:pos} we get from \Cref{s-prop:ipw} that  
    \begin{equation*}
        \N 
        = \E \left\{ \frac{\sum_{j=1}^k \#\mY_j}{k G_{\bullet \bullet}(w)} \right\}. 
    \end{equation*}
    The result follows by plugging in the form of $G_{\bullet\bullet}$ in \Cref{eq:gdd-simp}, that relies on \Cref{asp:repr,asp:unif}. 
\end{proof}

\begin{proof}[Proof of \Cref{thm:identif2}] 
    Applying \Cref{s-thm:identif2}, which holds under \Cref{asp:pos,asp:repr2,asp:perf-detect}, we get  
    \begin{equation*}
        \N
        = \E \left\{ \frac{f^*(0) \sum_{j=1}^k \#\mY_j}{k \,\d \Pi_{\bullet\bullet}(0)} \right\}. 
    \end{equation*}
    The result follows by plugging in the form of $\Pi_{\bullet\bullet} = \frac{1}{kn} \sum_{i=1}^n \sum_{j=1}^k \Pi_{ij}$ in \Cref{eq:pi-simp}, that directly relies on \Cref{asp:unif} and secondarily relies on \Cref{asp:repr} to remove the dependence on $i$ from the coverage probability.  
\end{proof}

\begin{proof}[Proof of \Cref{lem:f}]
    This follows immediately from \Cref{s-lem:f}. 
\end{proof}

\begin{proof}[Proof of \Cref{thm:identif}]
    Applying \Cref{s-thm:identif}, which holds under \Cref{asp:pos,asp:repr2,asp:unif2,asp:perf-detect}, we get  
    \begin{equation*}
        \N
        = \E \left\{ \frac{f(0) \sum_{j=1}^k \#\mY_j}{k \,\d \Pi_{\bullet\bullet}(0)} \right\}. 
    \end{equation*}
    The result follows by plugging in the form of $\Pi_{\bullet\bullet} = \frac{1}{kn} \sum_{i=1}^n \sum_{j=1}^k \Pi_{ij}$ in \Cref{eq:pi-simp}, that relies on \Cref{asp:unif}, as    \begin{equation*}
        k\, \d \Pi_{\bullet\bullet}(0)
        = \frac{1}{n} \sum_{j=1}^k \sum_{i=1}^n \frac{\P\{x_i \in a(\mS_j; w)\}}{w} 
        = \frac{1}{w} \sum_{j=1}^k \P\{x \in a(\mS_j; w)\},  
    \end{equation*}
    where the last equality applied \Cref{asp:repr}. 
\end{proof}

\section{Proofs for \texorpdfstring{\Cref{sec:example}}{Section example}}

\begin{proof}[Proof of \Cref{prop:no-vert}]
    It is clear that $S_1$ must have a Lebesgue density, say $h$.
    
    Consider any two points $u_1, u_2 \in [y,1-y]$ such that $|u_1 - u_2| \leq 2w$. Then the expression
    \begin{equation*}
        \P \left\{ \left|\frac{u_1+u_2}{2} - S_1 \right| \leq \frac{|u_1 - u_2|}{2} \right\}
        = \int_{\left[ e_1 - \frac{|u_1 - u_2|}{2}, e_1 + \frac{|u_1 - u_2|}{2} \right]} \, h(u)\,\d u
    \end{equation*}
    does not depend on $e_1=\frac{u_1+u_2}{2}$, by assumption, so its derivative with respect to $e_1$ vanishes. That is, 
    \begin{equation}
        h(u_2) - h(u_1)
        = 0,
    \end{equation}
    show that the density of $S_1$ is equal at $u_1$ and $u_2$. 
    
    It is clear that for any two points $u_1,u_2 \in (0,1)$, there exists a sequence of points $u_3, \dots, u_m$, for some integer $m$, such that $u_1 < u_3 < \cdots < u_m < u_2$ and each neighboring point (e.g. $u_1$ and $u_3$) satisfy the premise of the above construction. Thus $h(u_1) = h(u_2)$, and $h$ is uniform over $(0,1)$. 
\end{proof}

\begin{proof}[Proof of \Cref{prop:luigi2}]
    For any $e_1 \in [0,1]$, the coverage probability $\P ( |e_1 - S_1| \leq y ) = 2y / (1+2w)$ for any $y \in [0,w]$. This shows \Cref{asp:unif} holds by considering $e_1 = x_1, \dots, x_{\N}$. 
\end{proof}

\section{Proofs for \texorpdfstring{\Cref{sec:est}}{estimation section}}

\begin{lem}
\label{s-lem:f-correct}
    If $F_{\no} = F_{\no}(\cdot; \bar\theta)$ for some $\bar\theta$, then $\bar\theta = \theta_0$. 
\end{lem}

\begin{proof}[Proof of \Cref{s-lem:f-correct}]
    The expression being maximized is 
    \begin{align*}
        \E \left\{ \sum_{j=1}^k \sum_{Y \in \mY_j} \log \f_{\no}(Y; \theta) \right\} 
        & = \E \left\{ \sum_{j=1}^k \sum_{i=1}^{\N} R_{ij} \log \f_{\no}(Y_{ij}; \theta) \right\} \\ 
        & \hspace{-15mm} = \int \log \f_{\no}(y; \theta) \,\d \left\{ \sum_{j=1}^k \sum_{i=1}^{\N} \P(R_{ij}=1) F_{ij}^*(y) \right\} \\ 
        & \hspace{-15mm} \propto \int \log \f_{\no}(y; \theta) \,\d \left\{ \frac{\sum_{j=1}^k \sum_{i=1}^{\N} \P(R_{ij}=1) F_{ij}^*(y)}{\sum_{j=1}^k \sum_{i=1}^{\N} \P(R_{ij}=1)} \right\} \\ 
        & \hspace{-15mm} = \int \log \f_{\no}(y; \theta) \,\f_{\no}(y; \bar\theta) \,\d y. 
    \end{align*}
    A standard argument now shows that the maximizer $\theta_0$ satisfies $\bar\theta=\theta_0$. 
\end{proof}

\begin{proof}[Proof of \Cref{thm:est-working}]
    It follows immediately from the product rule that the nonparametric efficient influence function of $\tilde\psi(\P)$ is 
    \begin{equation*}
        \frac{w}{\P \{ \x \in a(\mS; w) \}} \biggl[ \E \left( \# \mY \right) \mathrm{EIF} + \f_{\no}(0; \theta_0) \Bigl\{ \# \mY - \E \left( \# \mY \right) \Bigr\} \biggr], 
    \end{equation*}
    where $\mathrm{EIF}$ denotes the nonparametric efficient influence function of $\f_{\no}(0; \theta_0)$. 
    
    It follows from standard M-estimation theory \citep[cf.][Theorem 5.23]{van1998asymptotic} that the maximum working likelihood estimator 
    \begin{equation}
        \arg\max_{\theta} \E_n \left\{ \sum_{j=1}^k \sum_{Y \in \mY_j} \log \f_{\no}(Y; \theta) \right\}, 
    \end{equation}
    is regular and asymptotically linear with influence function 
    \begin{equation*}
        - V(\theta_0)^{-1} \sum_{Y \in \mY_j} \frac{\dot{\f}(Y; \theta_0)}{\f(Y; \theta_0)}. 
    \end{equation*}
    Since influence functions are unique in nonparametric models, this must be the nonparametric efficient influence function of $\theta_0$. It now follows from the chain rule that the influence function $\mathrm{EIF}$ of $\f_{\no}(0; \theta_0)$ is 
    \begin{equation*}
        \dot{\f}(0; \theta_0)^{\top} \left\{ - V(\theta_0)^{-1} \sum_{Y \in \mY_j} \frac{\dot{\f}(Y; \theta_0)}{\f(Y; \theta_0)} \right\}. \qedhere 
    \end{equation*}
\end{proof}

\begin{proof}[Proof of \Cref{prop:est-review}]
    Applying the M-estimation expansion in the proof of \Cref{thm:est-working} and a first-order expansion of the random denominator, we get
    \begin{align*}
        \hat\psi_k^{\mathrm{std}}-\tilde\psi
        & = \frac{1}{k}\sum_{j=1}^k \varphi(\mY_j;\PP)
        -\frac{\tilde\psi}{\P\{\x\in a(\mS;w)\}}
        \frac{1}{k}\sum_{j=1}^k
        \frac{\lambda\{a(\mS_j;w)\}}{\lambda(\mathcal{E})}
         +\tilde\psi+o_{\PP}(k^{-1/2}).
    \end{align*}
    The expectation calculation preceding the proposition now gives 
    \begin{equation*}
        h(\mS)
        = \frac{\tilde\psi}{\P\{\x\in a(\mS;w)\}}
        \frac{\lambda\{a(\mS;w)\}}{\lambda(\mathcal{E})}.
    \end{equation*}
    When the known denominator is used, the expansion contains no transect-only term, so $h=0$.
\end{proof}

\begin{proof}[Proof of \Cref{thm:est}]
    By the law of large numbers and consistency of the plugin influence function, the definitions of 
    $\widehat\beta_k$ and $\beta_t$ give $\widehat\beta_k-\beta_t = o_{\PP}(1)$. 
    Since $\E_0\{t(\mS)\}=0$, we similarly get $\frac{1}{k}\sum_{j=1}^k t(\mS_j) = O_{\PP}(k^{-1/2})$. 
    Thus we get the asymptotic equivalence 
    \begin{equation*}
        \frac{1}{k}\sum_{j=1}^k
        \left\{\widehat m_k(\mS_j)-m_t(\mS_j)\right\}
        = o_{\PP}(k^{-1/2}).
    \end{equation*}
    Applying the M-estimation expansion in the proof of \Cref{thm:est-working} and plugging this conclusion 
    into the definition of $\widehat\psi_k$ gives
    \begin{equation*}
        \widehat\psi_k-\tilde\psi(\PP)
        = \frac{1}{k}\sum_{j=1}^k
        \left\{\varphi(\mY_j;\PP)-m_t(\mS_j)\right\}
        +o_{\PP}(k^{-1/2}),
    \end{equation*}
    as desired.
\end{proof}

\end{document}